\documentclass[sigconf]{acmart}
\usepackage{amsmath,amsthm,amsfonts,amssymb,amscd}
\usepackage{enumerate}
\usepackage{mathrsfs}
\usepackage{xcolor}
\usepackage{graphicx}
\usepackage[utf8]{inputenc}
\graphicspath{{figures/}}
\usepackage{pdflscape}

\usepackage{hyperref}

\usepackage{listings}
\usepackage{fancybox}
\usepackage{hyperref}
\usepackage{rotating}
\usepackage{adjustbox}
\usepackage{pdflscape}

\usepackage{array,etoolbox}
\preto\tabular{\setcounter{magicrownumbers}{0}}
\newcounter{magicrownumbers}
\def\rownumber{}

\usepackage[ textsize=tiny, textwidth=1.4cm]{todonotes}

\newcommand{\Part}[1]{\noindent\textbf{#1}}

\newlength\lunderset
\newlength\rulethick
\newcommand{\Space}[1]{}

\usepackage{xspace}

\newcommand{\Comment}[1]{}
\newcounter{observation}
\newcommand{\observation}[1]{\refstepcounter{observation}
        \begin{center}
        \Ovalbox{
        \begin{minipage}{0.93\columnwidth}
                \textbf{Observation \arabic{observation}:} #1
        \end{minipage}
        }
        \end{center}
}

\title{Examining User Reviews of Conversational Systems} 
\subtitle{A Case Study of Alexa Skills}

\author{Soodeh Atefi}
\affiliation{%
  \institution{University of Houston}
  \city{Houston, TX}
  \country{USA}}

\author{Andrew Truelove} 
\affiliation{%
  \institution{University of California at Irvine}
  \city{Irvine, CA}
  \country{USA}}
\author{Matheus Rheinschmitt}
\affiliation{%
  \institution{Federal University of Bahia}
  \city{Salvador, BA}
  \country{Brazil}}
\author{Eduardo Almeida}
\affiliation{%
  \institution{Federal University of Bahia}
  \city{Salvador, BA}
  \country{Brazil}}
\author{Iftekhar Ahmed}
\affiliation{%
  \institution{University of California at Irvine}
  \city{Irvine, CA}
  \country{USA}}
  
\author{Mohammad Amin Alipour}
\affiliation{%
  \institution{University of Houston}
  \city{Houston, TX}
  \country{USA}}
\begin{document}
\begin{abstract}
Conversational systems use spoken language to interact with their users.
Although conversational systems, such as Amazon Alexa, are becoming common and can provide interesting functionalities, there is little known about the issues users of these systems face.

In this paper, we study user reviews of more than 2,800 Alexa skills to understand the characteristics of the reviews and the issues that they raise. Our results suggest that most skills receive fewer than 50 reviews. Our qualitative study of user reviews using open coding resulted in identifying 16 types of issues in the user reviews. 
Issues related to content, integration with online services and devices, errors, and regression are the top issues raised by the users. Our results also indicate differences in volume and types of complaints by users when compared with more traditional mobile applications. We discuss the implication of our results for practitioners and researchers.



\end{abstract}
\maketitle

\section{Introduction}
Conversational systems that use spoken language to interact with a computing system have become steadily popular in recent years.
In addition to mobile phones and computers, nowadays there are dedicated devices, such as Amazon Echo, that  provide conversational interfaces through which spoken language is the  \emph{only} way to interact with the user. Presently, millions of such devices are being used by end users; as of January 2019, more than 100 million Alexa-enabled devices are in use\footnote{
\url{https://www.theverge.com/2019/1/4/18168565/amazon-alexa-devices-how-many-sold-number-100-million-dave-limp}}.

The growing popularity of conversational systems has entailed exciting opportunities for developing novel services for users and for reaching out to new audiences.
Developers can use the infrastructure provided by companies, such as Amazon and Google, to develop new conversational systems. These conversational systems can be invoked and used through the corresponding devices. There are also web portals~\cite{skillStore,assistantStore} that provide information about the available conversational systems and allow users to leave reviews about them.

Despite such popularity, little is known about the issues that users of conversational systems face. 
Analyzing the user reviews would improve our understanding of such issues, and it would help practitioners and researchers to develop processes and techniques to improve users' satisfaction. 
More specifically, studying the issues in conversational systems would help us to understand the types of bugs and to develop testing techniques and practices to identify such issues in the development that can improve the overall quality of conversational systems.

User reviews have been used previously  to provide insights about 
the quality of computing systems, e.g., numerous studies on analysis of reviews in apps stores~\cite{khalid2015mobile}. Since conversational systems are relatively new and provide functionalities using a different interface than conventional systems, it is highly likely that reviews of conversational systems are going to have additional types of reviews and complaints along with previously identified types of reviews. However, to the best of our knowledge, there is no empirical study investigating the user reviews in conversational systems. 

In this paper, we study user reviews in Amazon Alexa conversational systems. These systems are called Alexa Skills, or skills for short. 
We analyzed user reviews of more than 2,800 Alexa skills, amounting to more than 100,000 reviews.
We used open coding ~\cite{fincher2005making,seaman2008defect,seaman1999qualitative} to find the common issues that users' complained about while using the skills. 

Our results suggest that Alexa skills receive relatively small number of user reviews (~90\% of skills received less than 50 reviews) compared to mobile apps (50\% of skills received less than 50 reviews) that can be due to various reasons, such as an inconvenient feedback process in the Alexa skill ecosystem.

We found 16 types of issues described in the user reviews, from which 9 are unique to conversational systems.
We found that, while the correctness of responses is important for user satisfaction, non-functional characteristics, such as quality and volume of voice, are also important to users.
This highlights the necessity of considering different aspects of a system, such as tone of voice and audio quality, while creating skills in order to ensure a fluent and natural conversation between the user and system. We also observed that a major portion of the user complaints are pertinent to using Alexa skills for connecting to and managing other devices and services. Moreover, we found that users experience a new form of regression in conversational systems. In this paper, we discuss the implication of our findings in testing and designing conversational systems. 



\Part{Contributions} This paper makes the following contributions.
\begin{itemize}
    \item We collected and analyzed the users reviews of more than 2,800 skills.
    \item We discuss the issues reported in the user reviews.
    \item We compare the issues in the Alexa skills and mobile apps. 
    \item We make the data available to public. 
\end{itemize}

\Part{Organization} Section ~\ref{sec:background} provides rudimentary information about Alexa skills.
Section ~\ref{sec:methodology} describes the methodology of this study. 
Section ~\ref{sec:results} describes the results of the study, and Section ~\ref{sec:discussion} discusses the results.
Section~\ref{sec:related} describes the related work. Section \ref{sec:threats} discusses threats to validity of the study, and Section \ref{sec:conclusion} concludes the paper.  

\section {Background}
\label{sec:background}
In this section, we describe the main concepts of Alexa conversational systems. 
The development of an Alexa skill consists of two main steps: (1) specifying conversation/dialog parameters, and (2) deploying the skill.

\subsection{Specifying the dialog}
Conversational systems interact with users through spoken language. 
Systems such as Amazon Alexa, Google Home and Apple Siri implement language understanding and generation capabilities that allow developers to
develop a conversational system solution;
in the Alexa ecosystem, such a conversational system solution is called an Alexa skill, or a skill, for short.

Amazon Alexa provides two main abstractions to facilitate development of a skill: \emph{intent} and \emph{entity (slot)}.
An \emph{intent} categorizes the intentions of a user in utterances,
and  \emph{entity} is a piece of information necessary for processing the user's intent.

Developers can use a JSON file to  specify intents and entities of a skill.
Figure ~\ref{fig:sample-specification} shows a snippet of specification of intents and entities in the \textit{High Low} skill\footnote{\url{https://github.com/alexa/skill-sample-nodejs-highlowgame}}, which is a game skill. 
In this Figure, the \texttt{invocationName} key defines the term that can be used to launch the skill. For this skill, a user can say, ``high low game'' to launch/activate the skill on her Alexa device. 
The \texttt{intents} key specifies the list of intents in the skill. In the example shown in the Figure, \texttt{NumberGuessInten} is the name of one of the intents. 
To identify an intent in utterances, developers are required to provide sample utterances for each intent. Values for the \texttt{sample} key specify sample utterances for each intent. 
In this example, "{number}", "is it {number}", "how about {number}", and "could be {number}" are sample utterances.
The items enclosed in curly braces represent the name of entities, the variables to be captured in utterances.

The entities or slots and their corresponding data types are specified by the \texttt{slot} key in the JSON file. 
In the example, it is expected that utterances contain a value for an entity named \texttt{number} that is of type \texttt{AMAZON.NUMBER}. \texttt{AMAZON.NUMBER} represents a numeral data type.  

\begin{figure}
\begin{footnotesize}
\begin{verbatim}
{
"interactionModel": {
"languageModel": {
"invocationName": "high low game",
"intents": [
...
{
"name": "NumberGuessIntent",
"samples": [
    "{number}",
    "is it {number}",
    "how about {number}",
    "could be {number}"
          ],
"slots": [{
    "name": "number", "type": "AMAZON.NUMBER"}]
...
}    
\end{verbatim}
\end{footnotesize}
\caption{Snippet of specification of a sample skill.}
\label{fig:sample-specification}
\end{figure}

\subsection{Deploying an Alexa skill}
An Alexa skill is started when the user expresses the invocation term and the device starts listening to the user's utterances.  
Alexa devices are increasingly attempting to perform as much language processing tasks as they can on the device before transferring the results to the Amazon Alexa server for further processing. 
If fulfilling the user's intents requires connection to other servers, the Amazon server would forward the messages from the device to those servers. 
Some skills, especially home automation skills, require communicating messages to and from other devices. 
In such cases, the message usually goes through multiple hops, including through the Amazon server, the device server, and finally to the intended external device. The messages from an external device to the Alexa device take the reverse route.


\section{Methodology}
\label{sec:methodology}
\begin{figure*}
    \centering
    \includegraphics[height=5cm]{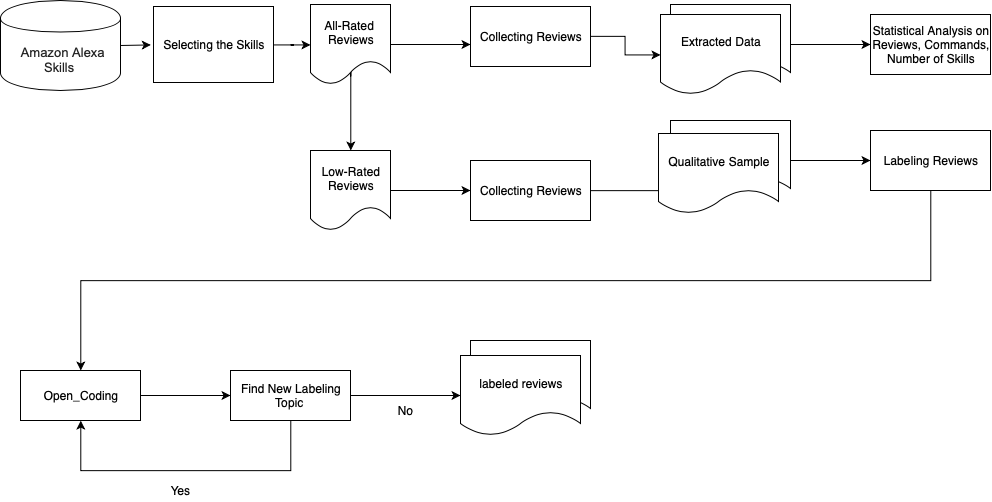}
    \caption{Overview of Our Approach}
    \label{fig:Workflow}
\end{figure*}

This section describes the methodology that was used to identify and evaluate the types of issues users write about in their Alexa skills reviews. Figure ~\ref{fig:Workflow} depicts the overall workflow in this study and the steps taken to analyze the reviews.

\subsection{Data Collection}
\begin{figure}
    \centering
    \includegraphics[width=\columnwidth]{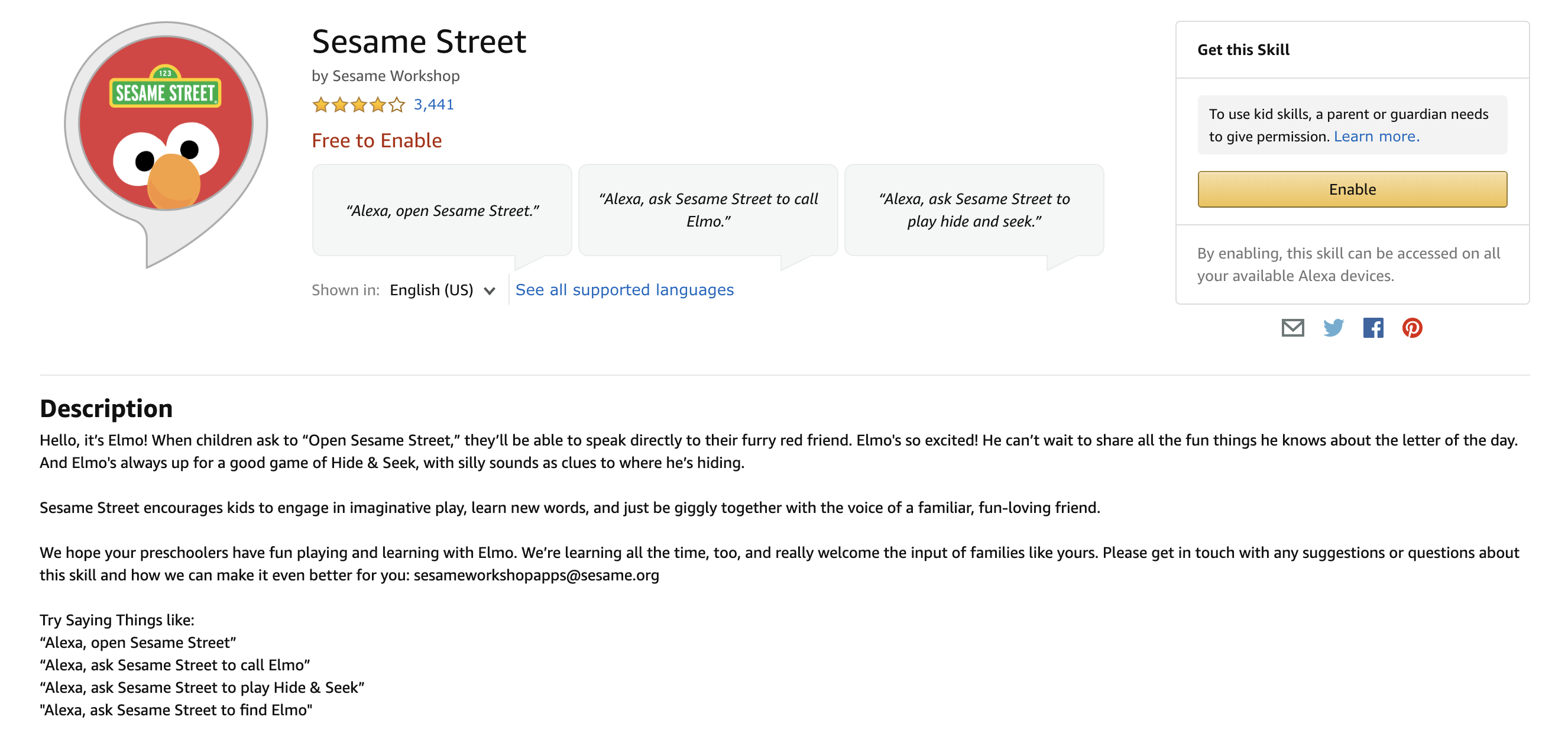}
    \caption{Sample skill with description}
    \label{fig:sesame-street}
\end{figure}

Every public Amazon Alexa skill has a web page on the Amazon website that contains information about the skill, including the skill name, the name of the publisher, a description of the skill's functionality,
and an option to enable or disable the skill on the devices that are connected to the user's Amazon account. In addition to basic information about the skill, the page provides examples of commands that can be used to interact with the skill. Figure ~\ref{fig:sesame-street} depicts an example web page for the \textit{Sesame Street} skill, where the statement, ``Alexa, ask Sesame Street to call Elmo'' is given as an example utterance. Amazon has organized the skills into 63 categories based on the skill's functionality, ranging from ``Accessories'' to ``Home Automation'' to ``Wine''. For referencer, the skill depicted in Figure~\ref{fig:sesame-street} belongs to the category of ``Games and Trivia''.


\begin{figure}
    \centering
    \includegraphics[width=\columnwidth]{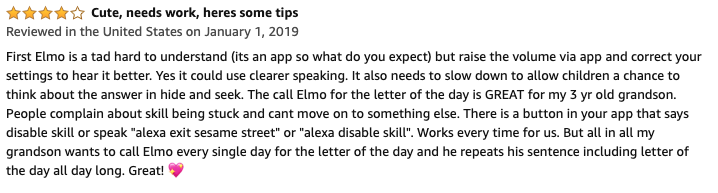}
    \includegraphics[width=\columnwidth]{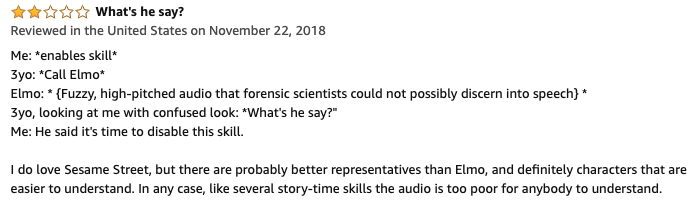}
    \caption{Sample reviews of skill in Figure~\ref{fig:sesame-street}}
    \label{fig:sample-reviews}
\end{figure}

\Part{Review Structure:}
Users with an Amazon account can post reviews for skills. Reviews encompass multiple components. Some of these components include an overall rating, a headline, and review text. An overall rating is a star-rating ranging from 1-star to 5-stars that indicates the user's satisfaction with the skill, 1 being minimal and 5 being maximum. A headline acts as the title for the review, and the text of the review contains the user's comments on the skill. Figure~\ref{fig:sample-reviews} shows samples of reviews. In addition to these components, each review includes data about the date the review was posted.

\Part{Data Collection:}
We used the \texttt{scrapy}~\footnote{\url{www.scrapy.org}} framework to develop a tool for extracting the information relating to the skills and their corresponding skill reviews.
For each skill category, our tool first extracts the URL of each skill's web page. It then collects all the reviews from the skill web pages. 
We did not impose any inclusion criteria on the skills, and included reviews as long as the Amazon server allowed us.  


\Part{Statistics of Collected Data: }
 Table \ref{tab:reviewspercategory} shows the total number of reviews per category, as well as the number of 1-star reviews and 2-star reviews. Since we wanted to gather an understanding about why users give low ratings, we focused on the ratings that were on the lower end of the star-rating system (1-star and 2-star), indicating an unsatisfactory experience using the skill. Overall, we collected more than 100,000 reviews from 2,817 skills. Smart Home and Streaming Services skills had the highest number of reviews. Kids and Currency Guides \& Converters skills had the lowest number of reviews among all categories.

\begin{table} 
\caption{Number of reviews per category}
\label{tab:reviewspercategory}
\resizebox{\columnwidth}{!}{%
\begin{tabular}{|c|c|c|c|}
\hline
\textbf{Category Name} & \textbf{\#Reviews} & \textbf{\#1-star} & \textbf{\#2-star} \\ \hline
Wine \& Beverages &89&20&7\\
Knowledge \& Trivia &4,867 &725&336\\
Home Services &634&209&35\\
Shopping &412&137&31\\
Game Info \& Accessories &522&63&23\\
TV Guides &75&35&7\\
Social Networking &43&6&1\\
Friends \& Family &180&58&15\\
Streaming Services &30,299&1,234&428\\
Movies \& TV &6 &2&2\\
Calendars \& Reminders &344&77& 22\\
Self Improvement &4,224&264&103\\
Translators &16&13&0\\
Public Transportation& 287&85&24\\
Connected Car &439&162&34\\
Food \& Drink &6&4&1\\
Kids &3&0&0\\
Taxi \& Ridesharing &5&2&0\\
Schools &29&1&1\\
Fashion \& Style &50&5&2\\
To-Do Lists \& Notes &255 &109&32\\
Pets \& Animals &1,750&276&103\\
Lifestyle &149&33&10\\
Device Tracking &155&47&8\\
Games &7,661&758&318\\
Navigation \& Trip Planners &330&100& 19\\
Games \& Trivia &1,315&152&48\\
Unit Converters &33 &13&1\\
News &4,344&1,389& 461\\
Currency Guides \& Converters &4&1&0\\
Music Info, Reviews \& Recognition Services &76& 19& 4\\
Podcasts &2,105&377&130\\
Cooking \& Recipes &1,257&348&111\\
Weather &1,924&538&169\\
Productivity &114&31&13\\
Flight Finders &169&92&17\\
Delivery \& Takeout &19&5&1\\
Restaurant Booking, Info \& Reviews &91 &40&8\\
Social &28&14&4\\
Organizers \& Assistants &1,435&338&102\\
Utilities &121&39&14\\
Event Finders &161&71&6\\
Travel \& Transportation &32&8&8\\
Zip Code Lookup &22&13&4\\
Dating &6&0&2\\
Astrology &134&45&18\\
Accessories &1,596&303&63\\
Score Keeping &69&11&9\\
Novelty \& Humor &2350 &630&115\\
Movie Info \& Reviews &43 &12&1\\
Alarms \& Clocks &352&122&28\\
Communication &1,103&267&50\\
Calculators &54 &22&5\\
Movie Showtimes &38 &21&0\\
Exercise \& Workout &33 &2&1\\
Education \& Reference &4,002&654&247\\
Business \& Finance &1105 &139 &39\\
Religion\& Spirituality &1,568&197&109\\
Local &10 &0&0\\
Music \& Audio &368 &79&20\\
Health \& Fitness &5376 &538&222\\
Smart Home &15,965 &7,130&1,457\\\hline
Total &100,252&18,085& 5,049\\ \hline

\end{tabular}%
}
\end{table}

\subsection{Qualitative Study}
To better understand the issues that users of Alexa skills face, we performed open coding \cite{fincher2005making,seaman2008defect,seaman1999qualitative} on reviews with low star ratings, following  prior studies in the domain of app review analysis, more specifically \cite{khalid2015mobile}. The coding process consists of two steps: code identification and open coding. 


\subsubsection{Code Identification}
\label{sec:code-identification}
We selected all 1-star and 2-star reviews available in the data set to create a pool of 23,134 low-rated reviews, i.e., 1-star and 2-star reviews.
We then randomly selected 100 of these reviews and used them in a preliminary coding process to help identify a set of codes that covered the range of different complaint types across the reviews
Four authors proceeded to code these reviews independently. 

The coding process for each author during this step consisted of examining reviews and identifying which code was most applicable in describing the primary type of complaint present in each review. In the event that a review contained a type of complaint that did not fall under any of the existing codes, the authors would create a new code that described the complaint type and added it to the list of codes. They would then restart the coding process from the beginning of the 100 reviews with the expanded list of codes.
The reason the coding would restart in this situation was to account for the possibility that a previously coded review might turn out to fit better under the new code than the code it had been assigned previously. 

For some reviews, we were unable to assign any specific code due to the vagueness of the complaint, e.g.,  "Don’t waste your time. Move on....". These reviews were coded as \textit{Not Specific}.

Once each of the authors finished independently coding the reviews, they discussed their resulting codes together, investigating the kinds of codes that each author had created. The authors also examined the places in which their codes aligned and where there was disagreement. They discussed the results and ultimately reached agreements on the final set of codes and their meanings. 16 different codes emerged from this process.
The discussion among the authors helped to eliminate any inconsistencies between the coded reviews.

\subsubsection{Open Coding}
\label{sec:open-coding}
Following the code identification, 
the authors used the final set of 16 codes to individually code reviews, basing their coding decisions on the discussion from the previous step. A total of 1,000 user reviews (500 1-star and 500 2-star reviews) were sampled and coded during this stage among the authors. During the process of coding these reviews, the authors continued to discuss and refine the codes in order to maintain a consensus on the way the reviews were coded. 

\subsection{Research Questions}
In this study we seek to answer the following research questions:

\begin{itemize}
    \item[RQ1] What are the characteristics of user reviews for the Alexa skills?
    \item[RQ2] What are the issues that users complain about?
    
    \item[RQ3] What types of complaints are unique to Alexa skills compared to the issues observed in iOS apps reported in the literature?
    \item[RQ4] What are the most frequent types of complaints?
\end{itemize}

\section{Results}
\label{sec:results}
\subsection{RQ1: User Review Characteristics}
\begin{figure}
    \centering
    \includegraphics[width=\columnwidth]{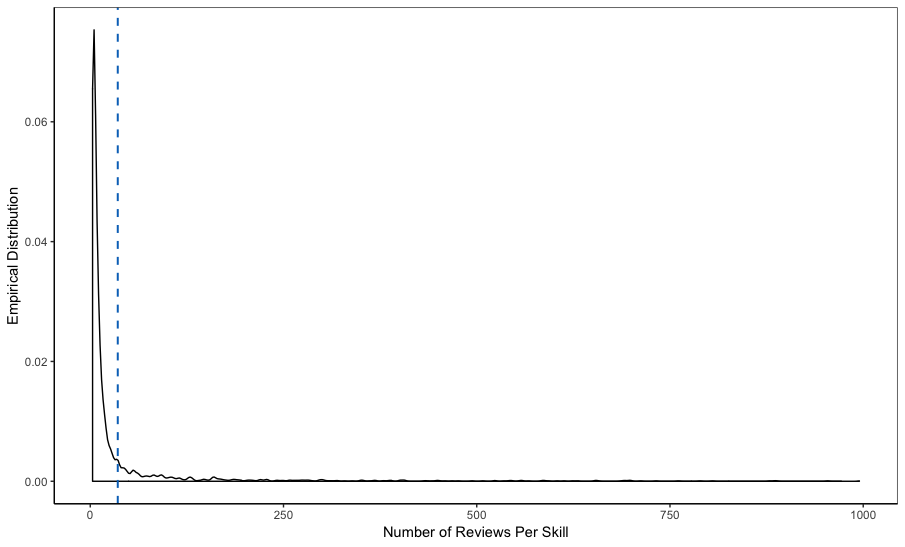}
    \caption{Empirical probability of number of reviews per app}
\label{fig:Distnumreviews}
\end{figure}

\Part{Number of Reviews Per Skill:}
Figure~\ref{fig:Distnumreviews} shows the empirical probability distribution of the number of reviews per skill. 
The area under the curve shows that most of the skills (87\%) have fewer than 50 reviews. 
``Streaming Services'', and ``Smart Home'' have the highest number of reviews among categories. 
There are a couple of possible reasons for the high number of reviews in these two categories. The first reason might be that most of these skills require users to use the skill web page in order to enable the services on their Alexa devices. Since they are already on the page, they might be more inclined to leave a review about their experience. 
The second possible reason could be related to the financial investment that users might have made in purchasing the subscriptions to their streaming services or home automation devices. This investment might provide an incentive to voice their opinions about the skill.

\observation{
$\sim$ 87\% of the skills have fewer than 50 reviews.}

\begin{figure}
  \centering
  \begin{minipage}[b]{0.4\textwidth}
    \includegraphics[width=\columnwidth]{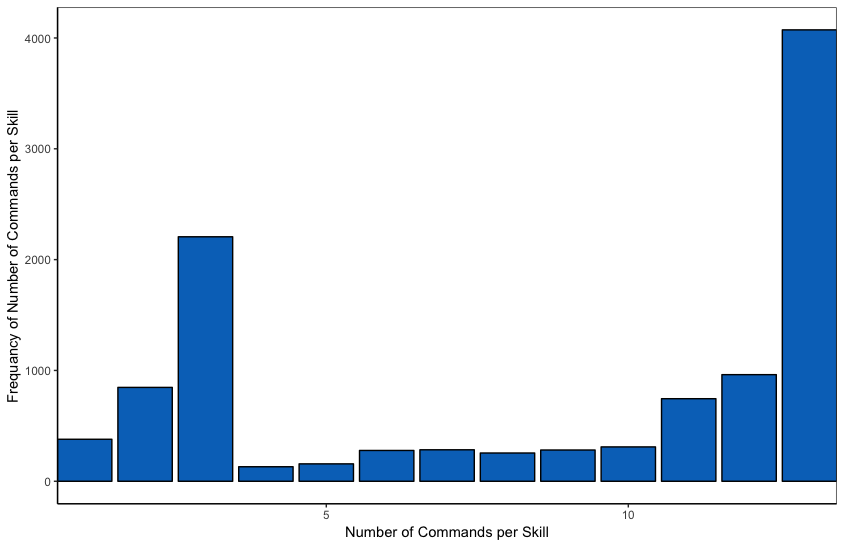}
    \caption{Number of Commands per Skill}
  \label{fig:commands_frequancy}
  \end{minipage}
  \hfill
  \begin{minipage}[b]{0.4\textwidth}
    \includegraphics[width=\columnwidth]{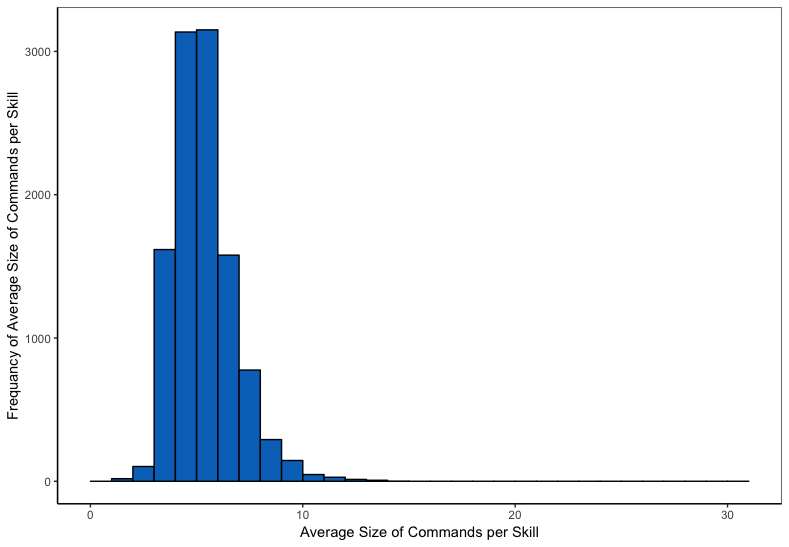}
    \caption{Histogram of Average Size of Commands in Skill}
 \label{fig:sizeCommands}
  \end{minipage}
\end{figure}

\Part{Number of Sample Commands:}
Figure \ref{fig:commands_frequancy} depicts frequencies for the 
different numbers of sample commands for the skills. The average number of sample commands per skill is 8.6. 25\% of the skills have three sample commands or fewer. Half of skills have 11 sample commands or fewer. A large portion of skills (37\%) have 13 sample commands.

\Part{Length of Commands:}
Each skill has a number of sample commands with different sizes (calculated in number of words).
Figure \ref{fig:sizeCommands} depicts the histogram of the average size of commands per skill.  
The average length of commands for the skills is 5.8 words. The average length of commands in half of the skills is more than six words.


\begin{figure}
    \centering
    \includegraphics[width=\columnwidth]{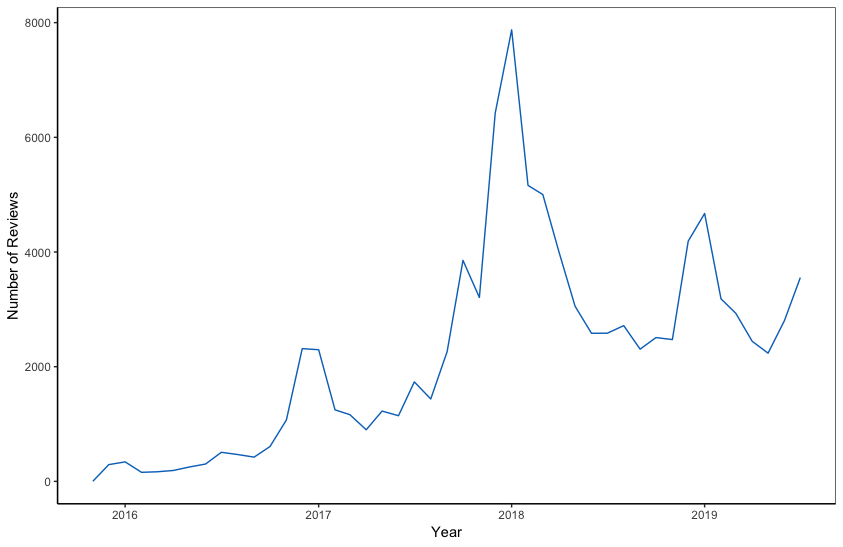}
\caption{Number of reviews over time}
\label{fig:time}
\end{figure}

\Part{Reviews per User and Temporal Distribution of Reviews:}
Looking at the authors of all reviews, 89\% of the users have written only one review, and 7\% of the users have written two reviews.
Figure~\ref{fig:time} shows the dates of the reviews, which covers 
the time period between 2017 and 2019. January of 2018 has the highest number of reviews, perhaps due to sales of Alexa devices related to the holiday season.

\subsection{RQ2: Issues and Their Frequency}

\begin{table*}
\caption{List of topics with definition and example}
\label{tab:results}
\resizebox{\textwidth}{!}{
\begin{tabular}{@{\makebox[3em]{\rownumber\space}}p{4cm}p{4cm}p{6cm}c}
\hline
\textbf{Name}
&
Brief Description
& Example &  Is New?  \\
\hline  
\gdef\rownumber{\stepcounter{magicrownumbers}\arabic{magicrownumbers}}\\
Content (Undesired Content, Uninteresting Content) & The system's content is  undesired or uninteresting & "I didn’t like the joke. they were dark jokes." &No\\ 
Error   & The system fails to perform correctly& "Alexa crashes after 3 questions" & No\\
Regression & The system does not work as it is used to (e.g., after update)& "Before the last update the bulbs were perfect now they don't work or just come off line.& No\\
Not Specific & A complain does not indicate a specific issue. & "It's a waste of time" & No \\
Feature Request &  A feature has been requested& "Please add the stock watchlist to the app so users can manually add/remove stocks from their list."& No\\
Ad & The system has in-app advertising or asked for subscription& "I tried to check for snow this week and had to wait through a long ad for premium." or "Its now subscription based. Disappointing." &No\\
Timing& The system has lags or the response time is inconsistent & "There's about a 5-10 second delay on this app (not on iphone)" & No\\\hline
Misunderstanding Intent& The system does not understand users' intent& "Alexa does nothing when asked to 'find' or 'ring my phone'."& Yes \\
Misunderstanding Entity (Slot)& The system does not understand the entities in the utterances& "ALEXA does not translate simple things like "translate kale in spanish.""& Yes\\
Dialogue Flow & Dialog is not fluent & "Alexa should just simply open and play the skill, without giving unwanted further advice or instructions."& Yes \\
Conversation Termination Error & User cannot end a conversations with a command & "I tried 'Alexa, cancel Stopwatch', 'Alexa, stop Stopwatch', 'Alexa, cancel all timers', but it would not stop the single allowed timer."& Yes\\
Audio Quality & Volume of the system is not appealing& "like everyone else says, the volume is too low compared to all the other news skills."&Yes\\
Commands &  Difficulties in the commands & "Too wordy. Need to simplify commands."& Yes\\
Integration with Devices & The system does not work properly with other devices& "Since adding Vera skill to Alexa I have not been able to add new devices to Alexa. The new devices linked to Vera but cannot be discovered by Alexa app."&Yes\\
Integration with Services & The system does not work properly with other services&"I tried for over an hour to get the skill to link to my Bose account."&Yes\\
Naturalness/ Too Mechanical
(Problem with Speech) & The system's voice is not appealing& "all I wanted to hear was Price Tag by Jessie J but then some robotic man started talking to me and I was really thrown off"&Yes\\
\hline
\end{tabular}%
}

\end{table*}
Table \ref{tab:results} shows the types of issues that were identified through the open coding. Within Table \ref{tab:results} are 16 codes corresponding to different complaint types, along with a brief description of the complaint type, and an example review to illustrate each issue. Column ``Is New?'' in the table signifies if similar issues have been observed before in the user reviews of mobile apps reported in \cite{khalid2015mobile}, or if the complaint type is new.
In this section, we first describe the complaint types that are comparable to the complaint types Khalid et al.\cite{khalid2015mobile} identified by mining user reviews of mobile apps. Then in the next subsection, we discuss the issues that are unique to Alexa skills. 

\Part{Content:} This issue relates to the content provided by the skill, such as news, stories, or jokes, that the user perceived as undesired or uninteresting. 
We found that 21\% of the reviews contain some issues pertinent to content, e.g., podcast or game. Below is an example of a content complaint where a user did not desire the provided content. 
\begin{quote}
    ``I thought this skill was a true question and answer application. Nope. It just tells you random facts''
\end{quote}

\Part{Error:} This issue points to instances in which a skill fails to perform correctly.
8\% of complaints were about this issue.

\Part{Regression:} This issue denotes cases in which users mention an adverse effect on the functional and non-functional properties of the skill following some update or change. 7\% of reviews contained ``Regression'' issues.

4\% of topics were ``Not Specific'' and ``Feature Request'' \cite{khalid2015mobile}
``Not specific'' means there was not a specific complaint in the content of the review that identified a particular issue.
In ``Feature Request'', users primarily requested a new feature or changes to a currently available feature. 

A small number of complaints (3\%) were about ``Money'' or ``Ad'' or were related to ``Timing''. In ``Money/Ad'', the system had an in-app advertisement (Ad) mechanism or asked users to make some kind of purchase, such as for a subscription. This issue type has been identified as ``Hidden Costs'' in iOS app user reviews. 

We found ``Timing'' can be comparable to ``Unresponsive Apps'' in iOS apps, which means that the service is slow to respond or function \cite{khalid2015mobile}.

\subsection{RQ3: Issues Unique to Conversational Systems}

\Part{Integration with Devices:} This issue refers to instances where a user reported some problem in using or managing a device, such as a thermostat or light bulb, via an Alexa skill. We found that 18\% of reviews contain issues pertinent to ``Integration with Devices''. The following is an example issue where the user cannot connect the skill to a device.
\begin{quote}
``It's a awesome camera ...
it won't work with Alexa. 
When given command I get a `mmm camera isn't responding'...''
\end{quote}

\Part{Integration with Services:} This issue denotes cases in which the skill is unable to properly work with an online service. We found 8\% of reviews mention this issue.
The following is an example of review that describes this type of issue.
\begin{quote}
``Can’t link the account. Registered an outlook address on the site - got the error message that `social accounts don’t work'.''
\end{quote}

Note that the issue of ``Integration with Services'' is different from the ``Compatibility'' issue that  Khalid et al.~\cite{khalid2015mobile} reported for mobile apps.
In their work, ``Compatibility'' denotes issues regarding the operation of an app on a specific device or operating system version.


\Part{Misunderstanding Intent and Entity:} 
This issue denotes cases where the user hints that the skill failed to understand some part of a user's utterances. We found that 8\% of topics in the reviews centered on a misunderstanding of utterances. More specifically, these cases involved a misunderstanding of either an utterance's intents or its entities. This lead to the creation of two codes: ``Misunderstanding Intent'' and ``Misunderstanding Entity''.
In the following example, the user complains that Alexa fails to  understand predefined commands. 
\begin{quote}
``failed over a dozen times, unable to get it to hear me correctly'' 
\end{quote}

\Part{Audio Quality:} This issue denotes reviews containing complaints about different characteristics of the audio in the skill, such as volume, or cadence of voice.
Around 4\% of reviews mentioned issues about audio quality.
For example, in the following review, the user complains that a skill in ``News'' category has poor sound quality.
\begin{quote}
``Great content but the sound quality is terrible.  I have to turn my volume way up in order to hear it, and if I am busy or forget to turn it down, I’m getting screamed at with further content from other skills.''
\end{quote}

We also found 6\% of user complaints 
were related to ``Commands'' or ``Dialog Flow''. 
In the following review, the user complains that a skill's commands are too long to use or remember. 
\begin{quote}
``Shorter \textbf{commands} would be much simpler to say and easier to remember''
\end{quote}

\observation {Users complain about skill commands that are long and not simple. According to our finding in Qualitative Results (4.1), the length of commands for most skills is greater than 4 words, with many over 8 words. These findings stress the importance of the command length that developers specify for their skills. Developers can also reduce the number of commands for a skill, which may make using a skill less complicated for users.}

There are also complaints about the flow of dialog. For example, in the following review, the user complains that the skill initiates unnecessary questioning.  
\begin{quote}
``This app would be great of it didn't ruin every "Goodnight" message with a follow up question asking me if I am interested in the developers' other apps and then Alexa stay on after the question and literally waits for a response...''. 
\end{quote}

Fewer than 5\% of reviews centered on ``Conversation Termination Error'' or ``Naturalness/Too Mechanical (Problem with Speech)''. 
In ``Conversation Termination Error'', complaints were related to the difficulty users experienced in terminating a conversation or skill with the specified commands.

\begin{quote}
``Me: *enables skill*\\
3yo: *Call Elmo* \\
Elmo: * {Fuzzy, high-pitched audio that forensic scientists could not possibly discern into speech} *\\
3yo, looking at me with confused look: *What's he say?"\\
...
I do love Sesame Street, but there are probably better representatives than Elmo, and definitely characters that are easier to understand.''
\end{quote}

It was also interesting to see that naturalness of the speech is not trivial for Alexa users.
For example, in the following review for a skill that reads the daily news, the user complains that the skill's voice is not natural and lacks human inflection.
\begin{quote}
``Definitely an interesting idea (which I like) but this skill has a lot of work to be done in order to be good. The intro is way too long and the voice sounds very robotic.''
\end{quote}

Table \ref{tab:complaint-frequency} depicts the most frequent 
complaint types.
Among the complaint types, 
``Content'', ``Integration with Devices'', ``Integration with Services'', ``Error'', and ``Regression'' occurred most frequently in the coded reviews.

\subsection{Most Frequent Complaints}
\begin{table} 
\caption{Most Frequent Complaint Types}
\label{tab:complaint-frequency}
\begin{tabular}{|c|c|}
\hline
Topics of Complaints &  Frequency\\
\hline
Content & 213\\ 
Integration with Devices & 189\\ 
Integration with apps/services & 83 \\
Error  &81\\
Regression & 75\\
Misunderstanding Intent & 50\\
Audio quality & 46\\
Not specific & 45\\
Feature Request & 45\\
Misunderstanding entity & 36\\
Command & 36\\
\hline
\end{tabular}

\end{table}

Table~\ref{tab:complaint-frequency} shows the number of reviews for the most prevalent issues.
More than 50 \% of complaints in the reviews fell under the labels of ``Content'', ``Integration with Devices'', ``Integration with Services'', or ``Error''.
We found that in our sample, most complaints under the ``Content'' complaint type were for skills in the ``News'' and ``Game'' categories. ``Integration with Devices'' and ``Integration with Services'' were  frequent complaints for skills in the ``Smart Home'' category.
Among specific complaints unique to conversational agents, ``Integration with Devices'' and ``Integration with Services'', ``Misunderstanding Intent'', ``Audio Quality'', ``Misunderstanding Entity (Slot)'', ``Commands'', and ``Dialogue Flow'' have the highest number of complaints. It can be nontrivial for developers and practitioners to pay attention to these issues when they are developing these systems that fall into the categories mentioned above. 
\section{Discussion}
\label{sec:discussion}
In this section, we discuss the results and their implications for practitioners and researchers. 
In particular, we interpret the results from two perspectives: a quality assurance perspective, and a software design perspective.

\subsection{Conversational Systems and Quality Assurance}
User reviews can draw attention to the  the aspects of a system that users value, and can impact the perceived quality of the system; consequently, they may impact the adoption and acceptance of the system.

\Part{Regression:}
We observed that some users complained about skills behaving differently after updates. It seems that in addition to regression in the main functionality of skill, users can be disappointed in changes to the content, audio quality, etc. The following example illustrates an example review where the user is upset about content and changes in volume. 

\begin{quote}
    I have been listening to Forest Night for almost a year now.  It WAS my absolute favorite! What happened over the weekend??  Where did the owl go??  Why is the volume so much louder??  This isn't an improvement!! Now Forest Night is just another cricket infested sound loop.  I'm so disappointed.
\end{quote}

This new type of regression poses interesting challenges to the software testing community relating to how to create test cases that can detect such changes to a skill. These issues suggest that the evolution of software for conversational systems is non-trivial, and developers should evaluate the ramifications of changes to the content and audio quality before they implement any modifications.

\Part{Commands and Quality:}
We observed that user reviews contained complaints about
long, wordy skill commands.
Simplicity of commands seems to be of importance to users of conversational systems, especially when longer utterances may increase the chance of a misunderstanding. 
Additionally, users may be displeased to discover that they will need to remember long commands in order to communicate with the skill. 
The need for simpler, more intuitive commands can be seen in the following review.
\begin{quote}
    ``I have Multiple Sclerosis ... My biggest issue with Alexa is that there are trigger words for apps. When I ordered Alexa with the hopes of it having abilities added like this, I had no idea I would have to remember trigger words for added capabilities especially in an emergency... Why is it not `Alexa I've fallen and I can't get up...' Or at least `Alexa I have an emergency...'''.
\end{quote}

\Part{Naturalness of Voice:}
Tone of voice can impact the perceived quality of the system. 
They add qualities to a conversation related to affect, such as emotion and sentiment, that are difficult to measure objectively and are varied based on personal and social experiences. A tone that sounds normal to one group of users can sound unnatural or belligerent to another group. For example, in the following review, the user perceived the voice in the skill to be unnatural.
\begin{quote}
``all I wanted to hear was Price Tag by Jessie J but then some robotic man started talking to me and I was really thrown off.''
\end{quote}
Utilizing human-centric guidelines~\cite{amershi2019guidelines} when designing conversational systems can address these issues to some extent.
The lack of objective metrics makes automating quality assurance for conversational systems challenging. 


\Part{Misunderstanding:}
Conversation is inherently prone to misunderstandings and often requires clarification and repetition of statements. One user complained about an instance of misunderstanding that had to be solved by altering enunciation in their utterances.
\begin{quote}
    ``100\% of the time Alexa thinks that I am talking about the LEXUS skill, not Linksys... 
I always have to really exaggerate the word Linksys in order to get the correct skill.''
\end{quote}

A skill that requires too much clarification or repetition from the user is undesirable and can adversely effect users' perceptions of the skill. This adverse reaction is reflected in the low-rated reviews which we coded as ``Misunderstanding Intent'' and ``Misunderstanding Entity (Slots)''. 

\subsection{Conversational Systems and System Design}
\Part{Comparison with the Issues Reported in Mobile App Reviews:}
We 
discovered that some of the complaint types we devised appear to be similar to the topics identified by Khalid et al. \cite{khalid2015mobile} for mobile app reviews. We identified 16 topics, out of which 7 matched with the topics identified by Khalid et al. \cite{khalid2015mobile}. This indicates that many of the identified topics are specific to conversational systems. 

\Part{Interaction Cardinality:}
Some users complained about situations where interaction with Alexa caused (or could have caused) embarrassment for them. For instance, in the following review, a skill's misunderstanding has caused ``disgust'' in the user.

\begin{quote}
...
``I was pretty curious to learn about a prehistoric fish named Dunkleosteus ...
"So I asked Alexa ask encyclopedia what is dunkleosteus". To my suprise and disgust Alexa ended up giving me the definition and description of what a "donkey show" was. Seriously what the heck ...
I would recommend anyone with children never to use this app to ask what a Dunkleosteus is...''
\end{quote}

Such issues reveal an interesting difference between traditional applications and conversational systems: \emph{cardinality}.
In traditional systems, the interaction is usually one-to-one; that is, a user that sits in front of the monitor, at the proper distance, can see the items on the screen. In conversational systems, however, the voice is unidirectional. 
Practitioners should be mindful of this characteristic and develop algorithms to avoid such situations. For example, the situation in the above comment could have been avoided if the skill had checked for sensitive content and sent a notification before beginning to read the Wikipedia page.

\Part{Presuppositions of Fluency of Conversation:}
Prior experience can impact users' perceptions and attitudes toward a system~\cite{assael1995consumer,karjaluoto2002factors}. 
While it likely that a system will be used by someone without any prior experience with similar systems, it is  unlikely that a user of an Alexa skill does not have a prior experience in conversation. Since early childhood, virtually all users have been reguarly engaging in conversations and have developed metrics over time to evaluate the fluency of conversations.  
This presupposition means that, for the success of a skill, understanding the target user population through user studies is necessary.

\Part{User Feedback:}
Mcilroy et al.~\cite{Mcilroy2017} reported that the median number of reviews for mobile apps is 50, but we found the around 90\% of skills receive fewer than 50 user reviews. Lack of user reviews for skills can hamper research in this area.  
One reason for the relatively lower number of user reviews for skills compared to mobile apps might be due to a more difficult feedback process. 

With mobile apps, it is easy to prompt users for a review and direct them to the app store to leave a review. With skills, users need to use the Alexa web page or mobile app to leave a review. This process is more cumbersome than with mobile apps, since there is no direct link from the skill to the review page. Devising methods to facilitate this process for conversational systems can be a promising line of research for improving the overall quality of these systems. 

\Part{Unclear System Boundary:}
In traditional software systems, the boundary of an application's active state is generally understandable by end users. An application is the combination of widgets on the screen that are active when the system is running and that disappear when the screen is closed. Users can usually answer questions about what applications are active on the system at any time. They can \emph{see} these applications, and they have access to tools such as Windows Task Manager to \emph{inspect} them.
With conversational agents, the notion of a skill is opaque to users. Users seem to be unsure about when a skill launches or when it terminates. 
Perhaps developers should make the points of entry and exit more explicit to the users.  

\Part{Conversation is Slow to Trigger:}
Interaction through conversational systems is not as flexible as with traditional display-based systems.
In traditional display-based human-computer interactions, the pace of interaction, in most cases, is determined by the speed of user's information processing. 
Users can pause, scroll up or down, or go back and forth when interacting with the screens on the display. Moreover, most screens allow users to terminate an application through a ``close'' button on the top of the screen.
This flexibility in navigation between and out of the screens can be easily adapted to an individual user's patterns of information processing and foraging. 
 
In contrast, conversational systems are less-adaptable to users' information processing characteristics. Alexa services allow users to use commands such as ``Alexa speak slower'' or ``Alexa speak faster'' to control the speed of conversation. User expectations about the ideal speed of conversation often varies depending on the circumstances, however. A user might likely find the need to utter such commands \emph{every} time that she wants to adjust the speed to be tedious.


\Part{Integration with Other Systems:}
Our results suggest that a significant number of reviews complain about integration with devices and online services. 
Users utilized Alexa skills to control and monitor various devices in their surrounding environment.
This functionality is realized through multiple levels of communication.
The Alexa device sends a message to the Alexa back-end server. The Alexa back-end server then sends messages to the server related to the device. The device server proceeds to send a message to the device; the response from the device then traverses the same route, but now in reverse order. There might also be other intermediate and integrative services such as \url{arlarm.com} and \url{ifttt.com} involved in this process. 
In this setting, any problem in the connections, servers, or messages can impede the transmission of data between an Alexa skill and the device. 
This multi-hop software system calls for development of tools and approaches to ensure the robustness of such systems.


\Part{Supportive community:}
Although we focused on relatively negative reviews, we observed that some users, like the one below, were also fairly supportive of the skill being reviewed, despite being unhappy about certain aspects of the skill. This group of users---critical, but supportive---can potentially be recruited judiciously to be involved in the development of a skill as beta testers of new features. 

\begin{quote}
``Ugh!!!! Microtranzations another game with great potential ... Overall the first chapter shows great potential and could be an amazing game. If it was free. However[,] the Star review might be a little too harsh. ...Overall not good but shows great potential''    
\end{quote}

\Part{Insufficient data for research:}
There are was one item of information that we wished was available on the pages of Alexa skills: the version of the skill.
Unfortunately, this page does not include any information about the current version or the release date of the skill. 
Having that information would have helped us to evaluate the possible changes in the attitudes of users with each new version.
\section{Related Work}
\label{sec:related}
Many researchers have analyzed user-reviews to detect user complaints for iOS and Android applications through the use open-coding. Khalid et al. \cite{khalid2015mobile} analyzed 6,390 1-star and 2-star rated user reviews of 20 iOS applications. They identified 12 different types of complaints in the reviews. Their results identified app crashes, feature requests, and functional errors as the most frequent complaint types. They also found that ethical issues, privacy, and hidden app costs had the most negative effect on the rating of an app. Hanyang Hu et al. \cite{hu2018studying} studied whether cross-platform mobile apps (apps that exist across different platforms) attain consistent star ratings and complaints across low-rated (1-star and 2-star) user reviews. They used open-coding to tag 9,902 low-rated reviews of 19 cross-platform apps. Their results showed that at least 68\% of cross-platform apps did not have a consistent distribution of star ratings on both platforms and that in 59\% of studied apps, complaints in the 1-star and 2-star reviews of the iOS versions of apps largely centered on app crashes.

Maalej and Nabil \cite{Maalej2015} proposed a probabilistic method to automatically classify app reviews and placed them into one of four categories: bug reports, feature requests, ratings, and user experience. They used review meta-data such as text classification, natural language processing, sentiment analysis, and simple string matching. 
They concluded that combining these techniques achieved better results than using any one of them separately.

Hoon et al. \cite{Hoon2013} studied the characteristics of reviews and how reviews evolve over time. 
Pagano and Maalej \cite{Pagano} explored information related to app reviews, such as how and when users leave feedback. The authors also analyzed the content of  the reviews.
AppEcho is a tool that allows users to add feedback in-situ when they face an issue~\cite{Seyff2014}.
AR-Miner is a framework for review mining that uses topic modeling to extract useful information from app reviews ~\cite{Chen2014}.

Guzman and Maalej ~\cite{Guzman2014} used natural language processing to extract fine-grained app features in the user reviews. 
Their process involved performing sentiment analysis to give each feature an overall sentiment score from the review. 
Topic modeling was then used to group features into more meaningful high-level features. 
Their results can help app analysts and developers quantify users' opinions when planning future releases.

Hermanson~\cite{Hermanson2014} examined whether the perceived ease of use and the perceived usefulness of an app were widely visible in user reviews on the Google Play Store. The author collected 13,099 reviews from the Google Play Store and discovered that only 3\% of the reviews contained information relating to perceived usefulness and that less than 1\% of the reviews had any mention of perceived ease of use.
The author’s results suggest that these qualities are not widely present in Google Play app reviews.

Panichella et al. \cite{Panichella2015} suggests using of Sentiment Analysis, Natural Language Processing, and Text Classification to classify the sentences in app reviews. They reasoned that deep analysis of the sentence structure can be exploited to find user’s true intentions rather than just topic analysis.
Truelove et al. \cite{truelove2019topics} extracted and analyzed Amazon product reviews for 10 different Internet of Things (IoT) enabled devices as well as the reviews from each device's corresponding mobile app from the Google Play Store; the analysis suggested that connectivity, timing, and updates were noteworthy topics of focus for developers of IoT systems.


Gu and Kim~\cite{Gu2016} proposed a tool called SUR-Miner for summarizing the reviews of apps.
The tool classifies reviews into five categories and finds the aspect of an app being discussed and the evaluation of the aspect, finally proposing outcomes in diagrams for the developer.
The authors analyzed the tool and found that it outperformed other methods in terms of accuracy. 88\% of developers that were surveyed in the study expressed satisfaction with the tool.
ARdoc is another tool designed to analyze app reviews~\cite{Panichella2016}.
It performs sentiment analysis, text analysis, and natural language parsing for classifying useful feedback in app reviews for software maintenance and evolution.

Khalid et al.~\cite{Khalid} examined the relationship between the results (error warnings) generated for an app by FindBugs---a static analysis tool--and the kinds of ratings and reviews the app received on the app store. They found that certain warnings from FindBugs such as ``Bad Practice'', ``Internationalization'', and ``Performance'' appeared more frequently in apps with low review scores. Additionally, they noticed that these warnings were reflected in 
the content of user reviews. 
They suggest developers should identify issues in FindBugs before releasing the app.


CLAP is a tool that help developers parse app reviews to helpdecide when to release an app update.~\cite{Villarroel}.
CLAP categorizes user reviews based on the content of the reviews.
It categorizes relevant reviews together, and then automatically prioritize categories the next app update.
Di Sorbo et al.~\cite{DiSorbo2016} propose a tool (called SURF) that summarizes app reviews and provides detailed information related to recommended updates and app changes.
ALERTme~\cite{Guzman2017} automatically classifies and ranks tweets on Twitter related to software applications using machine learning techniques. Evaluation of this tool shows that this tool has high accuracy.
Mujahid et al.~\cite{Mujahid2017} analyzed user reviews of wearable apps. They manually sampled and categorized six android wearable apps. Their results showed that the most frequent complaints were functional errors, lack of functionality, and cost.

There appears to be a lack of research focused on the types of user complaints that pertain to conversational systems, however. No research has been undertaken to investigate the various types of complaints users make with regards to interfacing with conversational systems like Alexa. 
With this study, we aim to fill this gap in the current body of research.
\section{Threats to Validity}
\label{sec:threats}
In this section, we describe several threats to validity for our study. We have taken care to ensure that our results are unbiased, and have tried to eliminate the effects of random noise, but it is possible that our mitigation strategies may not have been effective.

\Part{External Validity:} We collected more than 100,000 reviews from 2,817 skills and conducted an open-coding process on a sample of 1,000 user reviews (500 1-star and 500 2-star reviews). However, our samples were selected from only one source (Amazon Alexa skill reviews). Thus, our findings may be limited to skills available on the Alexa skill store. However, we believe that the large number of skills sampled from multiple categories more than adequately addresses this concern.

\Part{Internal Validity:} The manual coding of 100 reviews was done by five researchers. However, we believe that a sample of 1,000 reviews is a reasonable amount for having a good understanding about the types of complaints and the high inter-rater reliability among researchers during manual coding took care of this threat.
\section{Conclusion}
\label{sec:conclusion}
This paper presented the results of our analysis of user reviews of Alexa skills. 
We found 16 types of issues described in the user reviews, from which 9 are specific to conversational systems. 
We found that while the correctness of responses is important for user satisfaction, non-functional characteristics such as audio quality and volume of voice are also important to users.
This highlights that creating skills is not only a technical task; human aspects of designing a fluent conversation, such as tone of voice and audio quality, are important as well.
We also observed that many of the user complaints are pertinent to using Alexa skills for connecting to and managing other devices and services. Moreover, we found that users experience a new form of regression in conversational systems. 

Our work showcases that further research is needed to: a) understand the evolution and impact of reviews on skill quality, and b) build a support tool to help developers synthesize the reviews and prioritize their corrective effort accordingly.

\bibliographystyle{ACM-Reference-Format}
\bibliography{bib}
\end{document}